\documentclass[preprint,3p,number,a4paper,sort&compress]{elsarticle}
\usepackage{graphicx}
\usepackage{epstopdf}
\usepackage{dcolumn}
\usepackage{amsmath}
\usepackage{mathptmx, courier, pifont}
\usepackage{mathrsfs}
\usepackage[scaled=0.92]{helvet}
\usepackage[T1]{fontenc}
\usepackage{textcomp}
\usepackage{color}
\usepackage{lineno,hyperref}

\modulolinenumbers[1]

\journal{Nuclear Instruments and Methods in Physics Research Section B}

\begin{document}
\begin{frontmatter}

\title{An improvement of isochronous mass spectrometry: Velocity measurements using two time-of-flight detectors}

\author[USTC,IMP]{P.~Shuai}
\cortext[mycorrespondingauthor]{Corresponding authors}
\author[IMP]{X.~Xu\corref{mycorrespondingauthor}}
\ead{xuxing@impcas.ac.cn}
\author[IMP]{Y.~H.~Zhang\corref{mycorrespondingauthor}}
\ead{yhzhang@impcas.ac.cn}
\author[IMP]{H.~S.~Xu}
\author[IMP,GSI,Max-Planck]{Yu.~A. Litvinov}
\author[IMP]{M.~Wang}
\author[IMP,GSI,Max-Planck]{X.~L.~Tu}
\author[Max-Planck]{K.~Blaum}
\author[IMP]{X.~H.~Zhou}
\author[IMP]{Y.~J.~Yuan}
\author[IMP]{X.~L.~Yan}
\author[IMP,GSI,GUCAS]{X.~C.~Chen}
\author[IMP]{R.~J.~Chen}
\author[IMP,GUCAS]{C.~Y.~Fu}
\author[IMP,GUCAS]{Z.~Ge}
\author[IMP,GUCAS]{W.~J.~Huang}
\author[IMP,GUCAS]{Y.~M.~Xing}
\author[USTC,IMP]{Q.~Zeng}

\address[USTC]{Research Center for Hadron Physics, National Laboratory of Heavy Ion Accelerator Facility in Lanzhou and University of Science and Technology of China, Hefei 230026, People's Republic of China}
\address[IMP]{Key Laboratory of High Precision Nuclear Spectroscopy, Center for Nuclear Matter Science, Institute of Modern Physics, Chinese Academy of Sciences, Lanzhou 730000, People's Republic of China}
\address[GSI]{GSI Helmholtzzentrum f\"{u}r Schwerionenforschung, Planckstra{\ss}e 1, 64291 Darmstadt, Germany}
\address[Max-Planck]{Max-Planck-Institut f\"{u}r Kernphysik, Saupfercheckweg 1, 69117 Heidelberg, Germany}
\address[GUCAS]{Graduate University of Chinese Academy of Sciences, Beijing 100049, People's Republic of China}



\begin{abstract}
Isochronous mass spectrometry (IMS) in storage rings is a powerful tool for mass measurements of exotic nuclei with very short half-lives down to several tens of microseconds, using a multicomponent secondary beam separated in-flight without cooling.
However, the inevitable momentum spread of secondary ions limits the precision of nuclear masses determined by using IMS. 
Therefore, the momentum measurement in addition to the revolution period of stored ions is crucial to reduce
the influence of the momentum spread on the standard deviation of the revolution period, which would lead to a much improved mass resolving power of IMS.
One of the proposals to upgrade IMS is that the velocity of secondary ions could be directly measured by using two time-of-flight (double TOF) detectors installed in a straight section of a storage ring.
In this paper, we outline the principle of IMS with double TOF detectors and the method to correct the momentum spread of stored ions.

\end{abstract}

\begin{keyword}
\texttt ~Isochronous mass spectrometry\sep Storage ring\sep double TOF detectors \sep velocity measurement
\end{keyword}

\end{frontmatter}



\section{Introduction}

Isochronous mass spectrometry at heavy-ion storage rings plays an important role in mass measurements of short-lived nuclei far from the valley of $\beta$-stability~\cite{BLT}.
Many important results in nuclear physics and nuclear astrophysics have been obtained in recent years based on the accurate determination of mass values by applying IMS at Gesellschaft f\"{u}r Schwerionenforschung (GSI) and Institute of Modern Physics (IMP), Chinese Academy of Sciences~\cite{BLT,Xu}.
The basic principle of IMS describing the relationship between mass-over-charge ratio ($m/q$) and revolution period ($T$) can be expressed in first order approximation~\cite{Franzke-MSR2008}
\begin{equation}
\frac{d T}{T}
=\alpha_p \frac{d (m/q)}{m/q}-\left( 1-\frac{\gamma^2}{\gamma_t^2} \right)\frac{d v}{v}, \label{eq1}
\end{equation}
where $v$ is the ion velocity.
The momentum compaction factor $\alpha_p$ is defined as:
\begin{equation}
  \alpha_p= \frac{dC}{C}/\frac{dB\rho}{B\rho}. \label{alphap}
\end{equation}
It describes the ratio between the relative change of orbital length $C$ of an ion stored in the ring and the relative change of its magnetic rigidity $B\rho$. $\gamma_t$, defined as $\alpha_p \equiv 1/\gamma_t^2$, is the transition energy of a storage ring.
By setting the Lorentz factor $\gamma$ of one specific ion species to satisfy the isochronous condition $\gamma=\gamma_t$, the revolution period of the ions is only related to its mass-over-charge ratio $m/q$, and independent of their momentum spreads.
Obviously, this property can only be fulfilled within a small mass-over-charge region, which is called the isochronous window~\cite{Geissel}.
However, for other ion species beyond the isochronous window, or even for the ions within the isochronous window, the isochronous condition is not strictly fulfilled. As a result, the unavoidable momentum spread due to the nuclear reaction process will broaden the distribution of revolution period, and thus, lead to a reduction in mass resolving power.

To decrease the spread of revolution period, it is obvious that the magnetic rigidity spread of stored ions should be corrected for.
However, the magnetic rigidity of each ion can presently not be measured directly, especially for the ions with unknown mass-to-charge ratio, according to the definition of magnetic rigidity:
\begin{equation}
B\rho
=\frac{p}{q}
=\frac{m}{q}\beta\gamma c. \label{eq2}
\end{equation}
We note that, as shown in Eq.~(\ref{alphap}), ions with the same magnetic rigidity will move  around the same closed orbit, regardless of their $m/q$.
Therefore, the correction of magnetic rigidity of the stored ions can be established via the correction of  their corresponding orbit.

To realize this kind of correction, one recently-proposed method is to measure the ion's position in the dispersive arc section by using a cavity doublet, which consists of a position cavity to determine the ion position and a reference cavity to calibrate the signal power~\cite{Chen2015,Sanjari}.
However, the establishment of this method may strongly depend on the sensitivity of the transverse Schottky resonator.

\begin{figure}\centering
	\includegraphics[angle=0,width=8 cm]{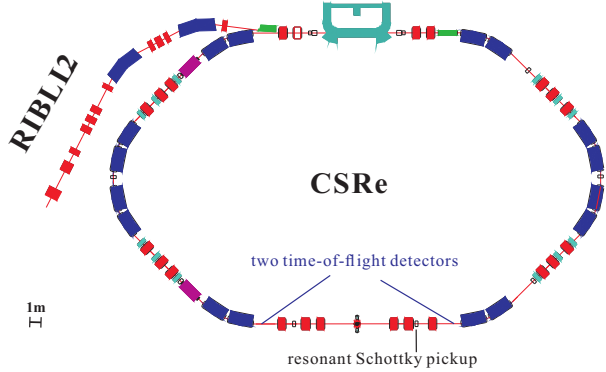}
	\caption{(Colour online) Schematic view of the arrangement of the double TOF detectors at the experimental cooler storage ring (CSRe) in IMP~\cite{Xia200211}.
		\label{Fig01}}
\end{figure}

Another possible method, which will be described in detail below, is to measure the ion velocity by using the combination of the two TOF detectors installed in a straight section of a storage ring. The original idea was first proposed at GSI~\cite{DoubleTOF}, and has recently been tested in on-line IMS experiments at IMP~\cite{XingYM} and studied by simulations~\cite{XuXing}. Fig.~\ref{Fig01} illustrates the schematic view of the setup of the double TOF detectors at the experimental cooler storage ring (CSRe).
By employing this additional velocity information of stored ions to correct the momentum spread, the mass resolving power of IMS will significantly be improved.

\section{Formulation and method}

In typical IMS experiments, the primary beam is accelerated and then bombards a target to produce the secondary particles.
After in-flight separation, several secondary ions are injected and stored in the ring simultaneously.
When the stored ions penetrates a thin carbon foil of the TOF detector, secondary electrons are released from the surface of the foil and are guided by perpendicularly arranged electrostatic and magnetic fields to a set of microchannel plate detectors (MCP).
The timestamps of each ion penetrating the TOF detector are recorded by a high-sampling-rate oscilloscope and are analyzed to determine the revolution periods of all stored ions. 
After correcting for instabilities of magnetic fields of the storage ring, all revolution periods of stored ions are superimposed to a revolution period spectrum, which can be used to determine the unknown masses and the corresponding errors~\cite{YHZhang,Tu, WSC}.

Obviously, the uncertainty of the revolution period, which directly relates to the error of the determined mass value, takes into account the contributions of momentum spread of stored ions.
In order to correct for such effects, we should think about the basic relationship of the revolution period ($T$) versus the velocity ($v$) of an individual ion, or the relationship of the orbital length ($C=T\cdot v$) versus the velocity.

In an IMS experiment, we can assume that when an ion was injected into the storage ring, its initial orbit and magnetic rigidity are $C_{init}$ and $B\rho_{init}$, respectively.
After penetrating the carbon foil of the TOF detector for many times, the ion's orbit and magnetic rigidity change to $C$ and $B\rho$, due to the energy loss in the carbon foil.
Because the relative energy loss after each penetration is tiny (in the order of $10^{-6}$), the change of the orbit can be regarded to be continuous.
Therefore, the relationship between the orbit and the magnetic rigidity can be obtained by integration of Eq.~(\ref{alphap}):

\begin{equation}
	\int_{C_{init}}^C \frac{dC}{C}
	=\int_{B\rho_{init}}^{B\rho}  \alpha_p \frac{dB\rho}{B\rho}. \label{eq3}
 \end{equation}


It is clear that the knowledge of $\alpha_{p}$ is crucial for solving this problem. In reality, the momentum compaction factor $\alpha_{p}$ is a function of the magnetic rigidity of the ring.
For an ion stored in the ring, it can be expressed as \cite{alphap}:
\begin{equation}
	\alpha_{p}\left( \beta\gamma\right) =\alpha_{p0}+\alpha_{p1}\frac{\delta \beta\gamma}{\beta\gamma}+\alpha_{p2}\left( \frac{\delta\beta\gamma}{\beta\gamma}\right) ^2+\cdots, \label{eq4}
\end{equation}
where $\alpha_{p0}$ is the constant part of the momentum compaction factor determined by the dispersion function, and $\alpha_{p1}$ is related to the perturbation of the momentum compaction factor, which has contributions from the slope of the dispersion function~\cite{alphap}. To clearly express the principle of IMS with double TOF detectors, firstly we ignore all the higher order components of $\alpha_p$ and consider the simplest approximation 
of $\alpha_p=\alpha_{p0}$.
The effect of higher order components of $\alpha_{p}$ will be discussed in the Section 4.

The result of the integration of Eq.~(\ref{eq3}) yields the relationship of the orbital length versus the velocity
\begin{equation}
	\frac{C}{(\beta\gamma)^{\alpha_{p0}}}
	=\frac{C_{init}}{(\beta_{init}\gamma_{init})^{\alpha_{p0}}}
	\equiv K. \label{eq6}
\end{equation}


We emphasize that $K$ is only determined by the kinetic parameters of a given ion, and keeps constant after the ion is injected into the ring, despite of its energy loss in the carbon foil of the TOF detectors.
According to Eq.~(\ref{eq6}), we can calculate the revolution period of each ion corresponding to any arbitrary orbital length, if we can measure the revolution period and the velocity of each ion simultaneously.
Therefore, by correcting the revolution period of each ion to a certain orbital length (equivalent to a certain magnetic rigidity for all ions), the corrected revolution periods can superimpose a spectrum with just higher-order contributions of momentum spread.
This is the cornerstone of our methodology.

Let us define a reference orbit with a certain length $C_0$, and the kinetic parameters of that ion circulating in this orbit are \{$T_0,~v_0,~\beta_0,~\gamma_0$\}. 
As discussed before, the reference orbit $C_0$ and the real orbit $C$ can be connected by the constant $K$:
\begin{equation}
\frac{C_0}{(\beta_0\gamma_0)^{\alpha_{p0}}}=
\frac{C}{(\beta\gamma)^{\alpha_{p0}}}
=K_{}. \label{eq8}
\end{equation}
Since the orbital length of a stored ion is: $C=T \cdot v$, the parameter $C$ can be determined experimentally.
The revolution period $T$ of the ion can be extracted from the timestamps of either of the two detectors using the previous method as described in Ref.~\cite{Tu}, and the velocity of the ion $v$ in any revolution can be directly measured by the double TOF detectors:
 \begin{equation}
 v
 =\frac{L}{t_{TOF1}-t_{TOF2}}, \label{eq9}
 \end{equation}
 where $L$ is the straight distance between the double TOF detectors, and $t_{TOF1},~t_{TOF2}$ are the timestamps recorded by them respectively~\cite{XuXing}.

\begin{figure}\centering
	\includegraphics[angle=0,width=8 cm]{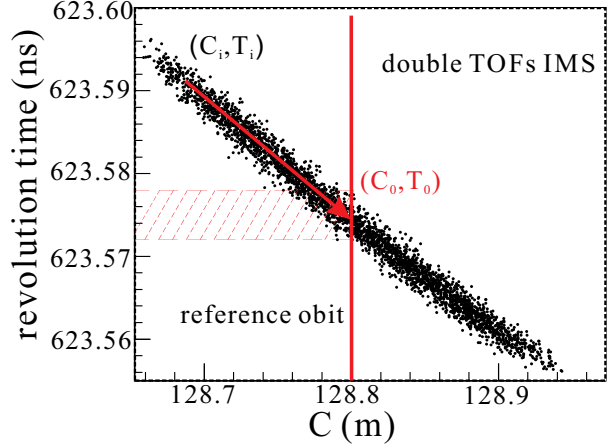}
	\caption{(Colour online) Simulated scatter plot (black points) of the averaged revolution periods versus the averaged orbital length for $^{25}${Al}$^{13+}$ ions~\cite{XuXing}. The red shadow shows the spread of revolution periods after orbital correction to a reference value at $C_0=128.801$ m.
		\label{Fig02}}
\end{figure}

In a real IMS experiment, a stored ion has betatron motions around the closed orbit, with the oscillation amplitude depending on its emittance.
During the time of acquisition in one injection (about $200-1000~\mu$s), the timestamps for one ion are recorded for hundreds of revolutions in the ring.
In consequence, the average values of $C$, $v$ and $T$ for each individual ion can be determined very precisely.

In principle, the reference orbit $C_0$ can be defined arbitrarily, even though the ions may never move on that reference orbit.
However, the central orbit of the storage ring is recommended in order to avoid the error caused by an extrapolation.
The momentum compaction factor $\alpha_{p}^{0}$ can be extracted from the lattice setting of the ring~\cite{GaoX} or be experimentally determined with a relative precision of $~10^{-3}$ by using the method described in~\cite{WSC}.

Finally, the velocity $v_0$ of the ion corresponding to the reference orbit $C_0$ can be calculated from Eq.~(\ref{eq8}), and then the revolution period $T_0= C_0/v_0$ is deduced. 
After repeating the procedure described above for each individual ion, all the obtained $T_0$ (corresponding to the same magnetic rigidity) can be accumulated into a revolution period spectra for mass calibration. 
In this way the resolution of the revolution period, and consequently the mass resolution, can be improved for all ion species. 

The relative precision of velocity measurements in IMS with double TOF detectors at CSRe can be roughly estimated.
Ions with $\gamma \approx 1.4$ will spend $\sim 86$~ns flying the $\sim 18$~m distance between the two TOF detectors in CSRe, and the time resolutions of both TOF detectors are $\sim 18$~ps (1 sigma) based on the results from an off-line test~\cite{Zhang20141}. So for an individual ions in one circulation, the relative precision of the time of flight, which is the same as the relative precision of velocity, is about $\sqrt{2}\times 18~ps/ 86~ns \approx 3\times 10^{-4}$.
Since the ions penetrate through the double TOF detectors for hundreds of times during the acquisition time, the fluctuations in the measurement of timestamps can be averaged in principle, leading to a relative precision of the velocity of $\sim 10^{-5}$, which is much better than the $B\rho$-acceptance of
about $\pm 0.2\%$ of CSRe~\cite{YHZhang}.

\section{Technical challenges}

\begin{figure*}\centering
	\includegraphics[angle=0,width=8 cm]{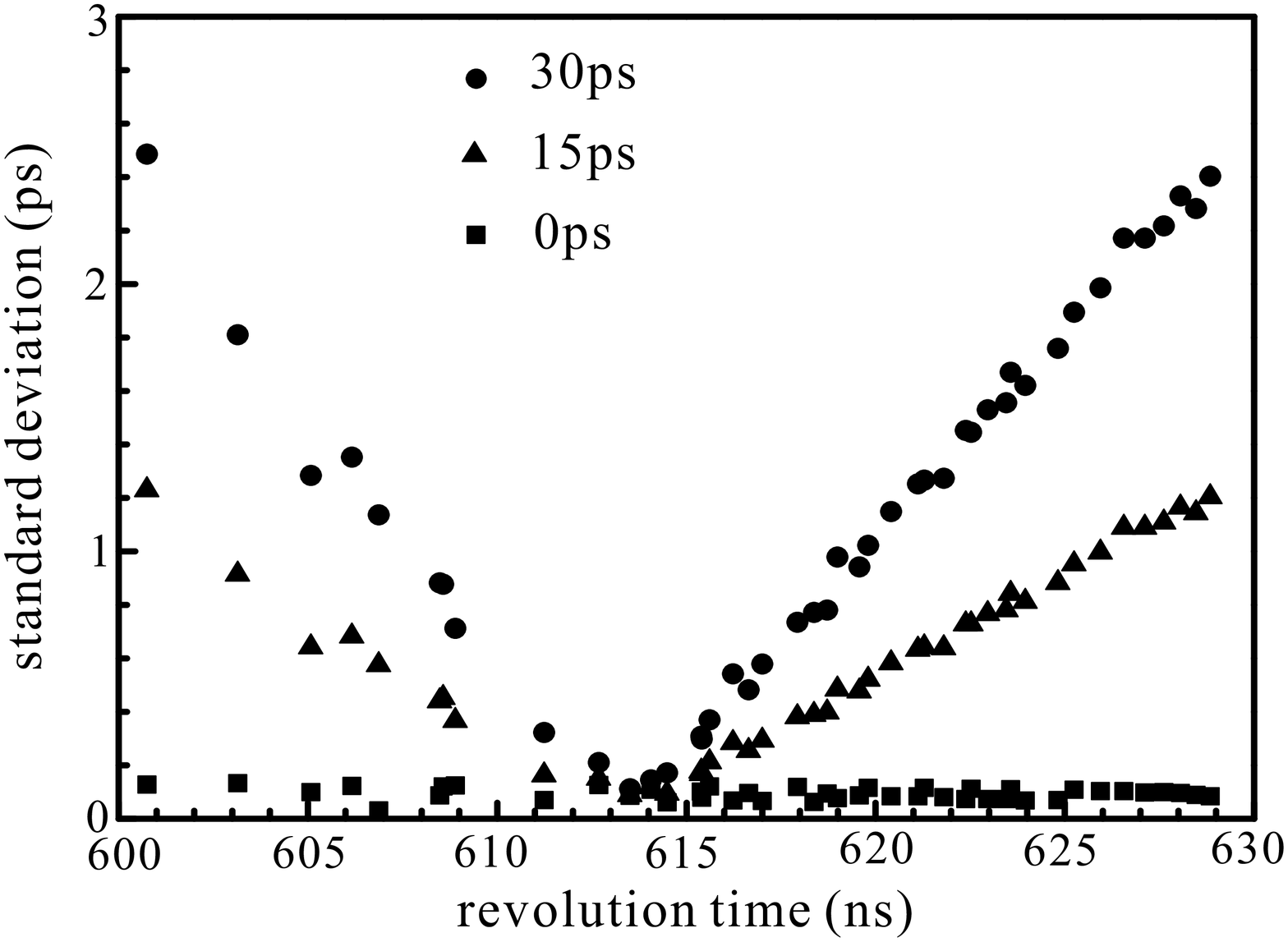}
	\includegraphics[angle=0,width=7.9 cm]{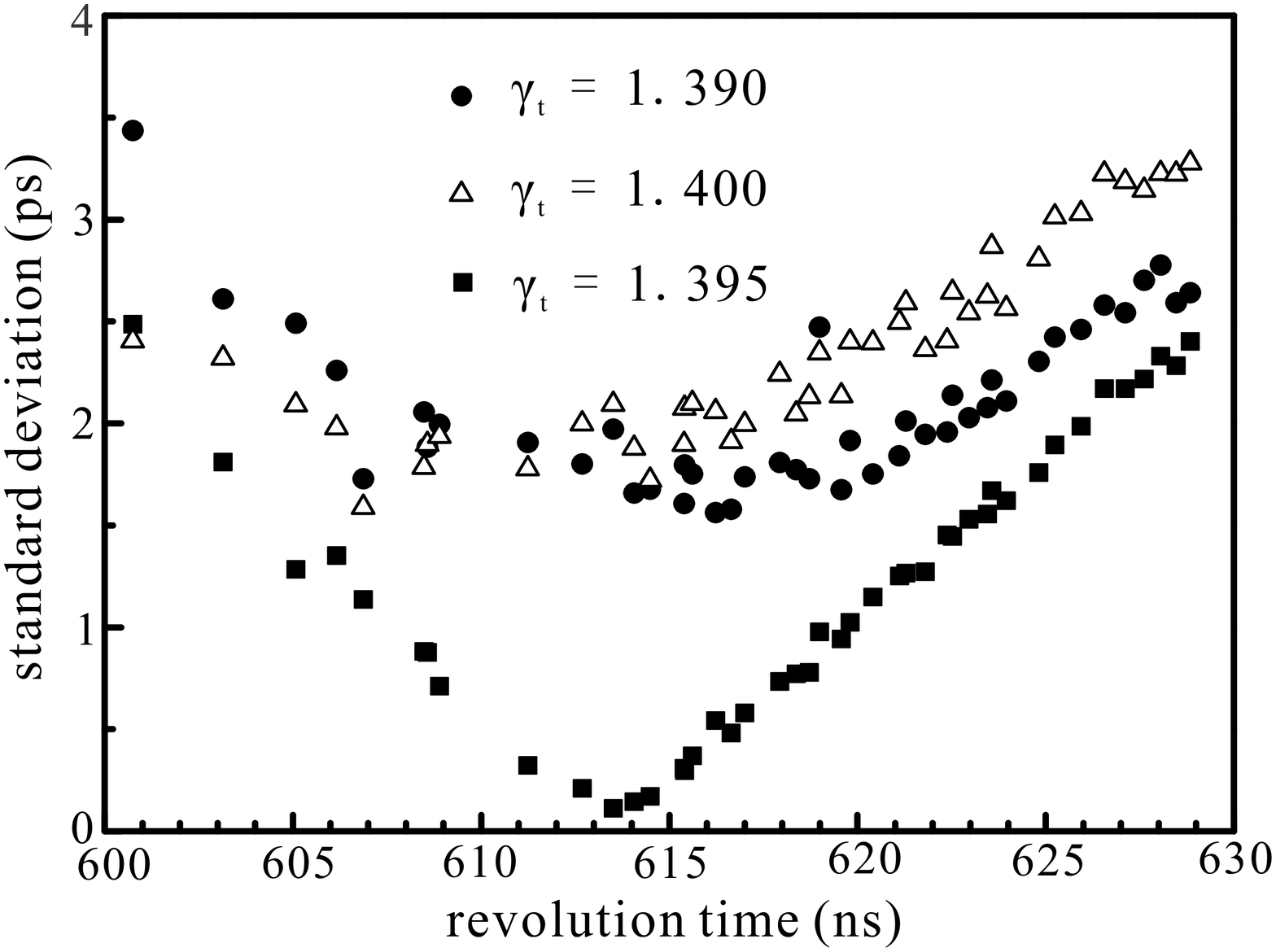}
	\caption{ (Left) Results from three simulations for IMS with double TOF detectors by setting the intrinsic time resolution of both TOF detectors to 30 ps, 15 ps, and 0 ps. (Right) Results from the same simulated data with real $\gamma_t=1.395$ and 30 ps time resolution of both TOF detectors but artificially applying three different $\gamma_t$ values in the data analysis. Solid circles, open triangles and solid squares represent results for $\gamma_t$ = 1.390, 1.400, and 1.395, respectively. Both pictures are taken from Ref~\cite{XuXing}.
		\label{Fig03}}
\end{figure*}

To test the principle of isochronous mass spectrometry with double TOF detectors, a Monte-Carlo simulation of an IMS experiment in the experimental storage ring CSRe has been made.
Six-dimensional phase-space linear transmission theory was employed to simulate the motions of stored ions in CSRe.
The timestamps of ions penetrating the double TOF detectors are generated by the simulation code, considering the beam emittance, the momentum spread of ions, and the energy loss in the carbon foil as well as the timing resolutions of the TOF detectors.
Then the method proposed in the above section can be applied to the simulated data.
The details of simulations can be found in Ref.~\cite{ChenRJ}.

Fig.~\ref{Fig02} illustrates the simulated results of the revolution period versus the orbital length of $^{25}${Al}$^{13+}$ ions.
The projection of the black points to the vertical axis represents the spread of the revolution period.
After transforming the revolution period at any orbit $C_i$ to the revolution period on the selected reference orbit $C_0$, the spread of revolution periods could be significantly improved (the area with red shadow).
However, to achieve a better mass resolving power, there are still some technical challenges to overcome.

It is clear that the spread of the corrected revolution periods of stored ions strongly depends on the precision of velocity measurement, which is limited by the timing performance of the TOF detectors according to Eq.~(\ref{eq9}) and so as the mass resolving power.
The left picture in Fig.~\ref{Fig03} illustrates the effect of the time resolution of the double TOF detectors on the revolution period resolving power.
Better time resolution of the TOF detector leads to higher mass resolving power, especially for the ion outside the isochronous window.
In the ideal case with no timing error, the standard deviation of revolution period approaches to zero,
and is almost the same for all ion species, which means the momentum spread of ions are no longer the main source of the revolution period spread.

In order to meet the requirement of a high-resolution TOF detector, two improved TOF detectors were tested offline and online at IMP in 2013~\cite{Zhang20141}.
By applying new cables with higher bandwidth and increasing the electric field strength and the corresponding magnetic field, the time resolution of the TOF detector in offline tests was significantly improved to 18 ps. 
However, TOF detector with higher time-resolving power is still highly recommended.

The determination of $\alpha_p$ is also very important to reduce the standard deviation of revolution periods.
The picture on the right in Fig.~\ref{Fig03} illustrates the impact of $\gamma_t$ on the corrected revolution period.
The mismatch between the real $\gamma_t$ of the storage ring and the parameter used in the data analysis will distort the systematic behavior of standard deviation as a function of the corrected revolution period.
This may be helpful for the determination of $\gamma_t$.

\section{Remarks}

In the discussion above, the momentum compaction factor is regarded for simplicity to be a constant.
In the real situation, the field imperfections of the dipole and quadrupole magnets will contribute to the high-order terms of the momentum compaction factor.
The nonlinear field is critical for the IMS experiment since it directly alter the orbital length variation and thus lead to a significantly lower mass resolving power~\cite{Dolin}. 
However, high-order isochronous condition can be fulfilled via some isochronicity correction by means of sextupoles and octupoles~\cite{GaoX,Dolin,Sergy}. 
Even though the momentum compaction factor can be corrected to be a constant within the $B\rho$-acceptance, we can still investigate the effect of $\alpha_{p1}$ on the relative variation of orbital length, which is the same magnitude as the relative variation of revolution period.
\begin{figure}\centering
	\includegraphics[angle=0,width=8 cm]{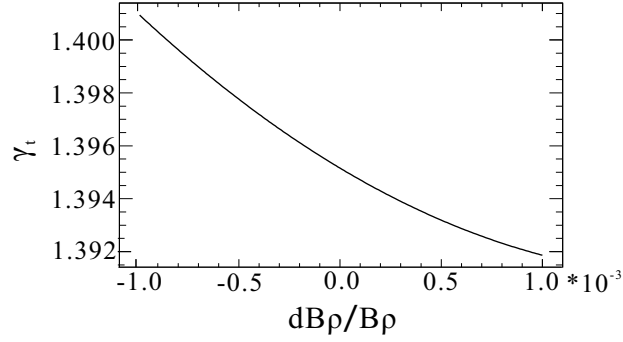}
	\caption{Dependence of the designed transition energy $\gamma_t$ of CSRe for the relative magnetic rigidities around the central magnetic rigidity. In the optical lattice of $\gamma_t=1.395$, the designed value of $\alpha_{p1}$ at CSRe is about $3$.
		\label{Fig04}}
\end{figure}

According to Eq.~(\ref{eq8}), the relationship between the momentum and the orbital length, for an individual ion, is independent from the selection of the initial orbit.
Therefore, in order to investigate the impact of $\alpha_{p1}$ on the ion orbit around the reference orbit $C_0$,
we can assume the initial orbit to be $C_0$, and a similar integration as Eq.~(\ref{eq3}) is obtained: 
\begin{equation}
\int_{C_{0}}^C \frac{dC}{C}
=\int_{\beta_{0}\gamma_{0}}^{\beta\gamma} ( \alpha_{p0}+\alpha_{p1}\frac{\Delta\beta\gamma}{\beta\gamma}) \frac{d\beta\gamma}{\beta\gamma}, \label{eq12}
\end{equation}
where $\Delta\beta\gamma=\beta\gamma-\beta_{0}\gamma_{0}$.
Simplifying the result of this integration, we can obtain:
\begin{equation}
	\frac{C}{(\beta\gamma)^{\alpha_{p0}}}
	\approx \frac{C_{0}}{(\beta_{0}\gamma_{0})^{\alpha_{p0}}}
	exp\left\{ \alpha_{p1}\left( \frac{\Delta\beta\gamma}{\beta_{0}\gamma_{0}}\right) ^2 \right\} . \label{eq13}
\end{equation}
The effect of $\alpha_{p1}$ on the orbital length can be estimated by comparing Eqs.~(\ref{eq6}) and  (\ref{eq13}) (with / without $\alpha_{p1}$).
The relative variation of the ion orbit is approximately determined by $\alpha_{p1}$ and the square of momentum spread $\delta p/p$ :
\begin{equation}
\frac{C_{wo}-C_w}{C_w}
\approx -\alpha_{p1}\left( \frac{\Delta\beta\gamma}{\beta_{0}\gamma_{0}}\right) ^2, \label{eq14}
\end{equation}
where $C_w$/$C_{wo}$ denotes the orbital length with/without the contribution of $\alpha_{p1}$.
As shown in Fig.~\ref{Fig04}, the designed value of $\alpha_{p1}$ at CSRe is about $3$ in the lattice setting with $\gamma_t=1.395$.
If the momentum spread $\delta p/p$ is at the magnitude of about $10^{-3}$, the contribution of $\alpha_{p1}$ on the relative change of orbital length would be about $3\times 10^{-6}$, which is the same magnitude of the mass resolution in present IMS~\cite{Tu, WSC}.

To eliminate the influence of $\alpha_{p1}$, we can simply define an effective orbital length $C_{eff}$:
\begin{equation}
	C_{eff}
	=C \cdot
	exp\left\{ -\alpha_{p1}\left( \frac{\Delta\beta\gamma}{\beta_{0}\gamma_{0}}\right) ^2 \right\}, \label{eq15}
\end{equation}
and then Eq.~(\ref{eq13}) becomes
\begin{equation}
\frac{C_{eff}}{(\beta\gamma)^{\alpha_{p0}}}
=\frac{C}{(\beta\gamma)^{\alpha_{p0}}} exp\left\{ -\alpha_{p1}\left( \frac{\Delta\beta\gamma}{\beta_{0}\gamma_{0}}\right) ^2 \right\}
=\frac{C_{0}}{(\beta_{0}\gamma_{0})^{\alpha_{p0}}}, \label{eq16}
\end{equation}
which is similar to Eq.~(\ref{eq6}).
Therefor, the constant $K$ for each ion can be determined using $C_{eff}$ instead of the real orbital length $C$.
In this way, the effect of the first-order term of momentum compaction factor is included, and can in principle be corrected.
However, to our knowledge, the important parameter $\alpha_{p1}$ can not be precisely measured in experiments.
It may be necessary to scan the $\alpha_{p1}$ as a free parameter in the data analysis until the minimum standard deviations of revolution period are achieved for all ion species.

\section{Summary}

In this paper, we present the idea of momentum correction for isochronous mass measurements in a storage ring by directly measuring the velocities of stored ions using the combination of two TOF detectors.
For all ions stored in the ring, their revolution periods can be corrected for using the information of ion velocity, so that all the corrected revolution periods correspond to the same reference orbit (or the same magnetic rigidity).
In this way, revolution periods of all ions with the same magnetic rigidity can be obtained, and thus leading to much improved mass resolving power for all ions. According to the results of simulations, the achievable mass resolving power of IMS with double TOF detectors strongly depends on the timing performance of the TOF detectors and the accuracy of the determination of $\gamma_t$.
Furthermore, the effects of high-order terms of the momentum compaction factor have been discussed, and besides the isochronicity correction by using sextupoles and octupoles, a possible solution to eliminate the influence of $\alpha_{p1}$ has been proposed.

\section*{Acknowledgments}

This work is supported in part by the 973 Program of China (No. 2013CB834401), National Nature Science Foundation of China (U1232208, U1432125, 11205205,11035007), the Chinese Academy of Sciences.
Y.A.L is supported by CAS visiting professorship for senior international scientists
and the Helmholtz-CAS Joint Research Group (Group No. HCJRG-108).
K.B. and Y.A.L. acknowledge support by the Nuclear Astrophysics Virtual Institute (NAVI) of the Helmholtz Association.


\end{document}